\documentclass{article}
\usepackage{ijcai26}

\usepackage{times}
\usepackage{soul}
\usepackage{url}
\usepackage[hidelinks]{hyperref}
\usepackage[utf8]{inputenc}
\usepackage[small]{caption}
\usepackage{graphicx}
\usepackage{amsmath}
\usepackage{amsthm}
\usepackage{booktabs}
\usepackage{algorithm}
\usepackage{algorithmic}
\usepackage{multirow}
\usepackage{makecell}
\usepackage{utfsym}
\usepackage{pifont}
\usepackage{xcolor}
\usepackage{colortbl, booktabs}

\usepackage[most]{tcolorbox}
\tcbuselibrary{theorems}
\newtcolorbox{promptbox}[1][]{
    colback=gray!5,
    colframe=gray!50,
    arc=3pt,
    boxrule=0.8pt,
    left=10pt, right=10pt, top=8pt, bottom=8pt,
    fonttitle=\bfseries,
    title=Prompt,
    #1
}

\title{Is AI Ready for Multimodal Hate Speech Detection? A Comprehensive Dataset and Benchmark Evaluation}

\author{
Rui Xing$^{1*}$\and
Qi Chai$^{2*}$\and
Jie Ma$^{1,3\dagger}$\and
Jing Tao$^1$\and
Pinghui Wang$^1$ \and
Shuming Zhang$^4$ \and
Xinping Wang$^3$ \and
Hao Wang$^2$ \\
\affiliations
$^1$MOE KLINNS Lab, Xi’an Jiaotong University\\
$^2$The Hong Kong University of Science and Technology (Guangzhou)\\
$^3$School of Cyber Science and Engineering, Xi’an Jiaotong University\\
$^4$Northwest University\\
\textsuperscript{*}Equal contribution\\
\textsuperscript{$\dagger$}Corresponding author\\
\emails
jiema@xjtu.edu.cn
}

\begin{document}

\maketitle

\begin{abstract}
Hate speech online targets individuals or groups based on identity attributes and spreads rapidly, posing serious social risks. Memes, which combine images and text, have emerged as a nuanced vehicle for disseminating hate speech, often relying on cultural knowledge for interpretation. However, existing multimodal hate speech datasets suffer from coarse-grained labeling and a lack of integration with surrounding discourse, leading to imprecise and incomplete assessments. To bridge this gap, we propose an \textbf{agentic annotation framework} that coordinates seven specialized agents to generate hierarchical labels and rationales. Based on this framework, we construct \textbf{M\textsuperscript{3}} (\textbf{M}ulti-platform, \textbf{M}ulti-lingual, and \textbf{M}ultimodal Meme), a dataset of 2,455 memes collected from X, 4chan, and Weibo, featuring fine-grained hate labels and human-verified rationales. Benchmarking state-of-the-art Multimodal Large Language Models reveals that these models struggle to effectively utilize surrounding post context, which often fails to improve or even degrades detection performance. Our finding highlights the challenges these models face in reasoning over memes embedded in real-world discourse and underscores the need for a context-aware multimodal architecture. Our dataset and code are available at \url{https://github.com/mira-ai-lab/M3}.

\textit{Disclaimer: This paper includes content that may be considered offensive or disturbing to some readers.}
\end{abstract}

\section{Introduction}
Hate speech~\cite{un_hate_speech_2019} refers to ``any kind of communication in speech, writing, or behavior that attacks or uses pejorative or discriminatory language with reference to a person or a group on the basis of who they are; in other words, based on their religion, ethnicity, nationality, race, color, descent, gender, or other identity factor.'' Its rapid dissemination across online platforms poses a serious threat to social stability~\cite{velasquez2021online}. For instance, the 2019 Christchurch mosque shootings in New Zealand and the terrorist attack in El Paso, Texas, that same year. Related reports and studies~\cite{barnes2019each,ware2022testament} indicate that the perpetrators incorporated Internet \textbf{Memes} into their manifestos or posts to spread hate speech and resonate with specific online subcultures, thus inciting extremist ideologies.

\begin{figure}[t]
\centering
\includegraphics[width=0.9\linewidth]{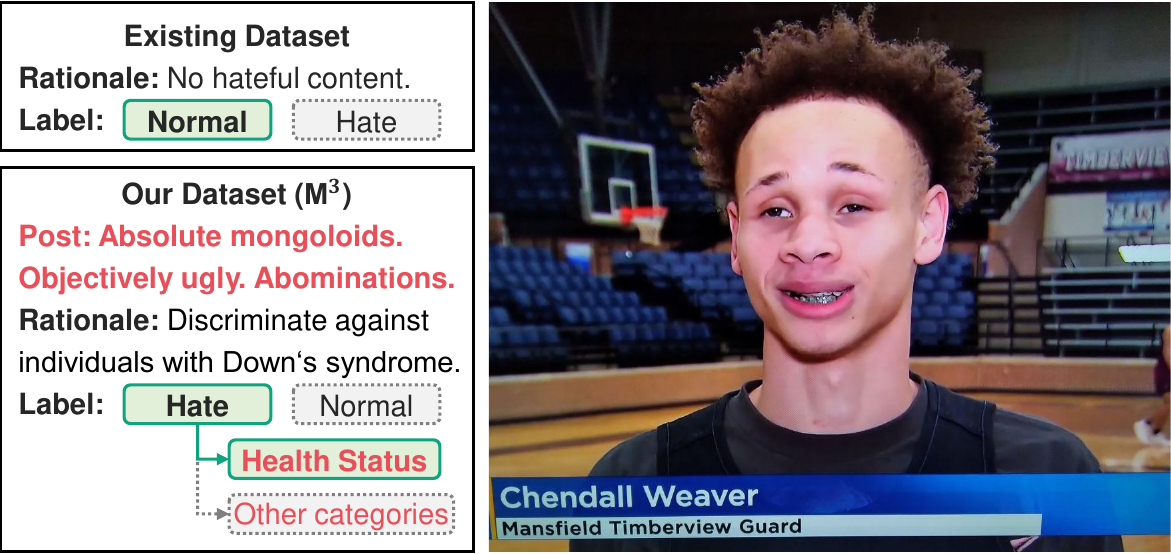} 
\caption{Comparison between existing datasets and ours (M\textsuperscript{3}). Existing datasets typically label the meme (right) as \textit{normal}. However, our dataset labels it as \textit{hate} with a refined classification of \textit{healthy state}-because of its accompanying post.}
\label{fig:example}
\end{figure}

\begin{table*}[t]
\centering
\renewcommand{\arraystretch}{1.2}
\resizebox{\textwidth}{!}{
\begin{tabular}{cccccccccc}
\toprule
\multirow{2.5}{*}{\textbf{Dataset}} & \multirow{2.5}{*}{\textbf{Domain}} & \multicolumn{2}{c}{\textbf{Label}} & \multirow{2.5}{*}{\textbf{Img text}} & \multirow{2.5}{*}{\textbf{Post}} & \multirow{2.5}{*}{\textbf{Rationale}} & \multirow{2.5}{*}{\textbf{Method}} & \multirow{2.5}{*}{\textbf{Source}} \\ \cmidrule(lr){3-4}
& & \textbf{Hateful?} & \textbf{Categories} \\ \midrule
\makecell{The Hateful Memes \\ Challenge Set} & Multiple fields & \makecell{Hateful, \\ Not-hateful} & - & \usym{2713} & - & - & Human-only & Synthetical \\ \midrule
HatReD & Multiple fields & \makecell{Hateful, \\ Not-hateful} & - & \usym{2713} & - & \usym{2713} & Human-only & Synthetical \\ \midrule
HarMeme & COVID-19 & \makecell{Very harmful, \\ Partially harmful, Harmless} & - & \usym{2713} & - & - & Human-only & \makecell{Google Image Search, \\ Reddit, Facebook, Instagram} \\ \midrule
MMHS150K & Multiple fields  & - & \makecell{Not hate, Racist, Sexist, \\ Homophobic, Religion, Other hate} & - & \usym{2713} & - & Human-only & Twitter \\ \midrule
MAMI & Misogyny & Misogynous, Not-misogynous & \makecell{Shaming, Stereotype, \\ Objectification, Violence} & \usym{2713} & - & - & Human-only & Twitter, Reddit \\ \midrule
ExMute & - & Hateful, Non-hateful & \makecell{Religious, Celebrity, \\ Political, Male, Female, Others} & \usym{2713} & - & - & Human-only & Facebook, Reddit, Instagram \\ \midrule
M\textsuperscript{3} (Ours) & Multiple fields & Hate, Normal & \makecell{Religion, Politics, Race, \\ Gender, Health Status, Violence, \\ Public Health, International Relations} & \usym{2713} & \usym{2713} & \usym{2713} & Human-validate agentic & \makecell{X, \\ 4chan, \\ Weibo} \\ \bottomrule
\end{tabular}
}
\caption{Comparison of representative multimodal hate speech datasets. While existing datasets often provide coarse labels or limited context, OURS introduces fine-grained multi-dimensional annotations and additionally includes the surrounding post content, enabling richer and more context-aware hate speech analysis.}
\label{tab:multimodal_datasets}
\end{table*}

A meme~\cite{shifman2013memes} is a multimodal composite consisting of an image and short text. As one of the vehicles for disseminating hate speech~\cite{pandiani2025toxic}, the meme is easily produced. It may combine a humorous image with a slogan that incites violence~\cite{zhou2021multimodal}, or convey discriminatory meanings through visual elements, with the embedded textual content appearing neutral in isolation~\cite{hee2024recent}. Within online communities, memes are often inconspicuously embedded in otherwise ordinary posts and are interpretable only by users who share specific cultural or subcultural knowledge. This subtle embedding—combined with their multimodal ambiguity—facilitates rapid dissemination through densely connected social networks~\cite{brown2018so,schmid2025humorous}. These characteristics not only make hateful memes particularly insidious but also expose critical gaps in current multimodal hate speech detection datasets.

Existing datasets~\cite{chhabra2023literature,jiang2024cross,nayak2022detection} predominantly support superficial evaluation, which fails to accurately assess the true performance of hateful meme detection methods, thereby offering limited constructive feedback for improving hate speech detection systems. \textit{On the one hand, they still adopt coarse-grained or flat labeling schemes}, creating a significant discrepancy with the multi-dimensional complexity of hate speech as defined by UN protocols. As illustrated in Figure \ref{fig:example}, the simplistic binary labels (i.e., Hate vs. Normal) preclude the representation of intersectional offenses, thus failing to reflect model efficacy in nuanced, multi-class scenarios. \textit{On the other hand, they typically isolate memes from their surrounding discourse}, focusing exclusively on their immediate content. Since memes in real-world social media are intrinsically tied to accompanying posts, the absence of such contextual information may lead to the erroneous label or even incomplete and misleading interpretations (see Figure \ref{fig:example}).

To address the challenges above, we propose an agentic annotation framework that coordinates seven specialized agents including a collector, extractor, cleaner, annotator, arbiter, explicator, and validator. The collector gathers memes and associated posts from X (formerly Twitter), 4chan, and Weibo, enabling broad cross-cultural coverage. After processing by the extractor and cleaner, the annotator and arbiter conduct multi-round hierarchical annotation with expert adjudication. Subsequently, the explicator generates hate rationales, while the validator performs quality control. This framework yields \textbf{M\textsuperscript{3}}, the \textbf{M}ulti-platform, \textbf{M}ulti-lingual, and \textbf{M}ultimodal \textbf{M}eme dataset, which contains 2,455 high-quality multimodal instances with two top-level labels (\textit{hate} and \textit{normal}), and eight fine-grained hate categories. We further evaluate M\textsuperscript{3} on state-of-the-art Multimodal Large Language Models (MLLMs), including Gemini-3, GPT-4o models and representative open-source models such as Qwen-VL, LLaVA-v1.6, and GLM-4.1V-9B-Thinking. Among them, Qwen3-VL-8B-Instruct achieves the highest overall score across the evaluation metrics. The results show that incorporating surrounding post context leads to degraded accuracy in hate detection and fine-grained classification. These findings suggest that current AI is not yet ready to fully replace human moderation. Moreover, hate speech detection should consider the linguistic context in which memes are embedded.

The main contributions are summarized as follows:
\begin{itemize}
\item We develop an agentic annotation framework, which coordinates multiple agents for scalable labeling and rationale generation, while ensuring annotation reliability through systematic human verification.
\item We introduce M\textsuperscript{3}, a multi-platform, multi-lingual, and multimodal meme dataset spanning X, 4chan, and Weibo, featuring fine-grained hate annotations and human-verified rationales to support explainable multimodal hate analysis.
\item We provide extensive benchmarking results on state-of-the-art MLLMs, demonstrating the effectiveness of M\textsuperscript{3} in evaluating multimodal hate meme detection capabilities and revealing current limitations in rationale generation.
\end{itemize}

\section{Related Work}

\subsection{Unimodal Hate Speech Datasets}
\label{sec:related-work}
Early research primarily focused on plain-text content via manual annotation (see Table \ref{tab:Unimodal_datasets} in Section \ref{sec:app-related-work} of the appendix. Initial English-centric efforts~\cite{Waseem2016hateful} were later scaled through crowdsourcing~\cite{founta2018large} and extended to non-English contexts, including Arabic~\cite{mulki2019hsab}, Chinese~\cite{rao2023Chinese}, and cross-lingual datasets~\cite{tonneau2025hateday}.

Parallel to this lingual diversification, the annotation paradigms shifted toward multifaceted schemes. OLID~\cite{zampieri2019predicting} employs a three-tier hierarchy, while HateXplain~\cite{mathew2021hatexplain} incorporated community-specific labels and rationales to enhance transparency. This explainability-oriented paradigm was later extended to non-English settings through HateBRXplain~\cite{salles2025hatebrxplain}.

Despite these advances, unimodal datasets remain limited in dynamic real-world scenarios. They strip away the visual and semiotic cues prevalent in social media platforms, creating a critical gap in capturing the full spectrum of hate speech.

\begin{figure*}[t]
\centering
\includegraphics[width=0.95\textwidth]{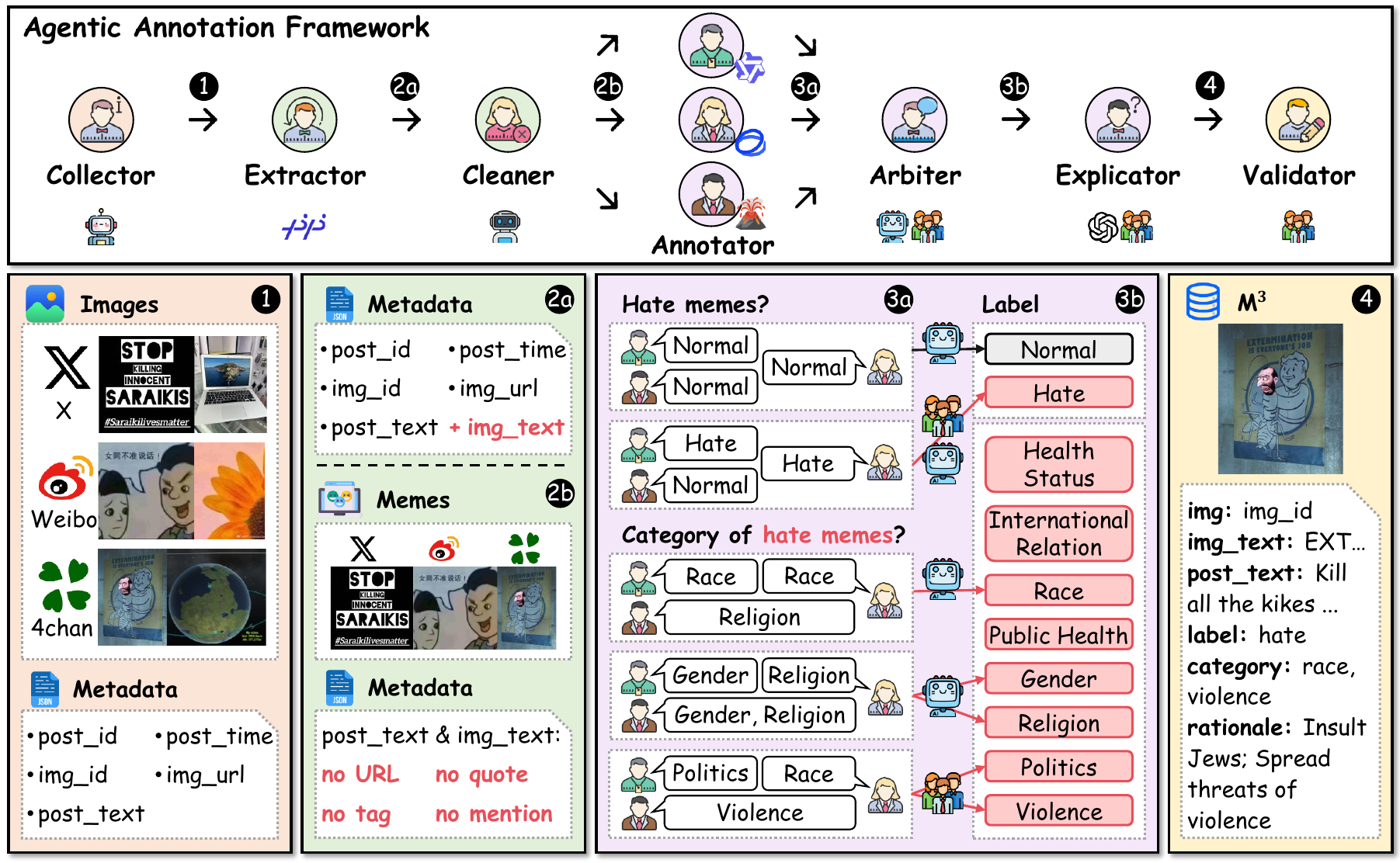}
\caption{The agentic annotation framework for M\textsuperscript{3}. \ding{182} Acquisition: Collector harvests multi-platform images and metadata. \ding{183} Preprocessing: Extractor and Cleaner perform OCR and metadata refinement. \ding{184} Annotation: Annotators, Arbiter, and Explicator collaborate on classification and rationale generation. \ding{185} Validation: Validator conducts quality assurance to finalize the M\textsuperscript{3} dataset (sample entry on the right).}
\label{fig:framework}
\end{figure*}

\subsection{Multimodal Hate Speech Datasets}

Existing datasets vary in data sources, annotation granularity, and target applications, as summarized in Table~\ref{tab:multimodal_datasets}. The Hateful Memes Challenge Set~\cite{kiela2020hateful} is among the earliest multimodal hate speech datasets, offering synthetically constructed samples with binary labels to assess the vision-language fusion capability of MLLMs. HatReD~\cite{hee2023decoding} further enriches annotations by capturing meme entities and additional socio-cultural context.

Efforts to capture real-world data, such as HarMeme~\cite{pramanick2021detecting} and MAMI~\cite{fersini2022semeval}, provide authentic samples beyond binary labels but are largely restricted to single-domain issues (e.g., COVID-19, misogyny). While MMHS150K~\cite{gomez2020exploring} incorporates post text, its labels remain relatively flat and lack the structured explanations.

Unlike most datasets that treat memes as isolated images, M\textsuperscript{3} preserves the multimodal context (memes + post) and provides hierarchical annotations that align more closely with complex real-world hate speech dynamics. Furthermore, unlike ExMute~\cite{debnath2025exmute}, which mainly focuses on Bengali and language-level phenomena, M\textsuperscript{3} provides broader multilingual coverage such as English and Chinese, bridging Western and Eastern social media ecosystems.

\section{Agentic Annotation Framework}
\label{sec:framework}
The Agentic Annotation Framework (illustrated in Figure \ref{fig:framework}) is a collaborative, multi-agent system designed to construct the \textbf{M\textsuperscript{3}} with hierarchical labels and rationales from raw social media data. It orchestrates seven specialized agents across four phases, forming a systematic pipeline for large-scale data acquisition and preprocessing, hierarchical annotation, and expert-driven quality assurance.

\subsection{Data Acquisition}
\paragraph{Collector.}
The workflow begins with the \textbf{Collector}, the primary gateway for data acquisition. Its input consists of platform-specific data streams, and it outputs images and raw metadata including \texttt{post\_id}, \texttt{post\_time}, \texttt{img\_id}, \texttt{img\_url}, and \texttt{post\_text} (\ding{182} in Figure \ref{fig:framework}). To capture the evolving nature of meme-based hate expression, the Collector continuously harvests data via APIs from X\footnote{\url{https://docs.x.com/x-api/introduction}}, Weibo\footnote{\url{https://open.weibo.com/}}, and the /pol/ (politically incorrect) board of 4chan\footnote{\url{https://github.com/4chan/4chan-API}}, spanning content from January to March 2024. In total, the Collector acquires 3,811,443 image–metadata pairs, including 43,567 from X, 2,090,793 from Weibo, and 1,677,083 from 4chan. Throughout the process, the Collector enforces privacy-preserving constraints by collecting only essential content without personal identifiers and securely storing all data in a local environment.

\subsection{Preprocessing}
\paragraph{Extractor.}
Following data acquisition, the \textbf{Extractor} initiates the processing phase by bridging visual and textual modalities in memes. Given the raw images gathered by the Collector, the Extractor invokes an OCR tool (PaddleOCR\footnote{\url{https://github.com/PaddlePaddle/PaddleOCR}} in our implementation) to extract text embedded within images (\texttt{img\_text}), which is then combined with the original post context (\texttt{post\_text}) to form a complete textual representation. Beyond recovering textual cues, the Extractor makes textual content observable to support strict downstream filtering, enabling subsequent agents to systematically identify images with excessively long text or no text at all. During this stage, some instances are inevitably filtered out due to unsuccessful tool execution.

\paragraph{Cleaner.}
The Cleaner subsequently refines the combined textual outputs (\texttt{post\_text} and \texttt{img\_text}) from the Collector and Extractor. It removes images whose embedding text is overly verbose or entirely absent, as well as samples containing platform-specific noise such as URLs, hashtags (\#), user mentions ($@$), and quoted content ($<<$). While images without textual content may still convey hateful intent through visual symbolism alone, enforcing such constraints is necessary to maintain high precision in large-scale meme filtering. Through text normalization and length-based filtering, the Cleaner produces a high-density multimodal candidate set of 535,471 samples (\ding{183} in Figure \ref{fig:framework}).

\subsection{Hierarchical Annotation}
\paragraph{Annotator.}
As an MLLM-driven expert, the three Annotators independently execute hate speech detection and category classification. Given the candidate memes, the Annotator performs coarse-to-fine annotation. It first conducts general hate speech detection to assign a binary Normal/Hate label. Then, conditioned on a hate decision, it carries out fine-grained classification across eight predefined domains (\ding{184} in Figure \ref{fig:framework}). The Annotator leverages the zero-shot capabilities of MLLMs to capture subtle, multilingual, and culturally contextualized hate expressions (implementation details in Section \ref{sec:app-annotation} of the appendix).

\paragraph{Arbiter.}
To ensure the reliability of these automated labels, the Arbiter acts as a consensus-monitoring unit. It aggregates multiple independent outputs produced by Annotators and evaluates their agreement at both the label and category levels. High-consistency predictions like unanimous hate labels or largely aligned category assignments (\ding{184} in Figure \ref{fig:framework}) are automatically accepted, while low-consistency samples are routed to a GUI (see Figure \ref{fig:hate-annotation} and Figure \ref{fig:category-annotation} in Section \ref{sec:app-annotation} of the appendix) for manual review. Samples with persistent ambiguities are excluded to maintain data integrity, leaving 3,179 high-quality verified annotations after this arbitration process.

\paragraph{Explicator.}
For samples identified as hate speech, the Explicator introduces an additional semantic layer by generating structured natural language rationales. Taking memes together with their accepted labels and categories as input, the Explicator synthesizes visual cues and textual context to articulate the rationale behind each classification. Concretely, its output follows a controlled verb–object phrase format, such as “mock a religious group” or “incite violence against immigrants”, which explicitly encodes the action and the targeted entity. 

\subsection{Quality Assurance}
\paragraph{Validator.}
Finally, the workflow concludes with the Validator (\ding{185} in Figure \ref{fig:framework}), which audits all candidate labels, categories, and rationales. This stage adopts voting without modification, in order to strictly assess the reliability of Arbiter's decisions. Three graduates with complementary disciplinary backgrounds (one in sociology and two in computer science) independently voted on 3,179 samples. Instances receiving only one vote are discarded, resulting in a final set of 2,455 samples. For a random audit of 200 high-consistency samples, inter-rater agreement remains high ($93.7\%$), further validating the effectiveness of the Arbiter in generating high-consistency annotations and stability of the proposed agentic annotation framework.

\section{M\textsuperscript{3}}

As the final outcome of our agentic annotation framework, Figure \ref{fig:framework} presents an example from the M\textsuperscript{3} dataset. M\textsuperscript{3} consists of several structured fields (\texttt{img}, \texttt{img\_text}, \texttt{post\_text}, \texttt{label}, \texttt{category}, and \texttt{rationale}), enabling a comprehensive assessment of MLLMs with respect to hateful meme detection, categorization, and explanation. We provide a detailed analysis of M\textsuperscript{3} below.

\begin{figure}[t]
\centering
\includegraphics[width=0.9\linewidth]{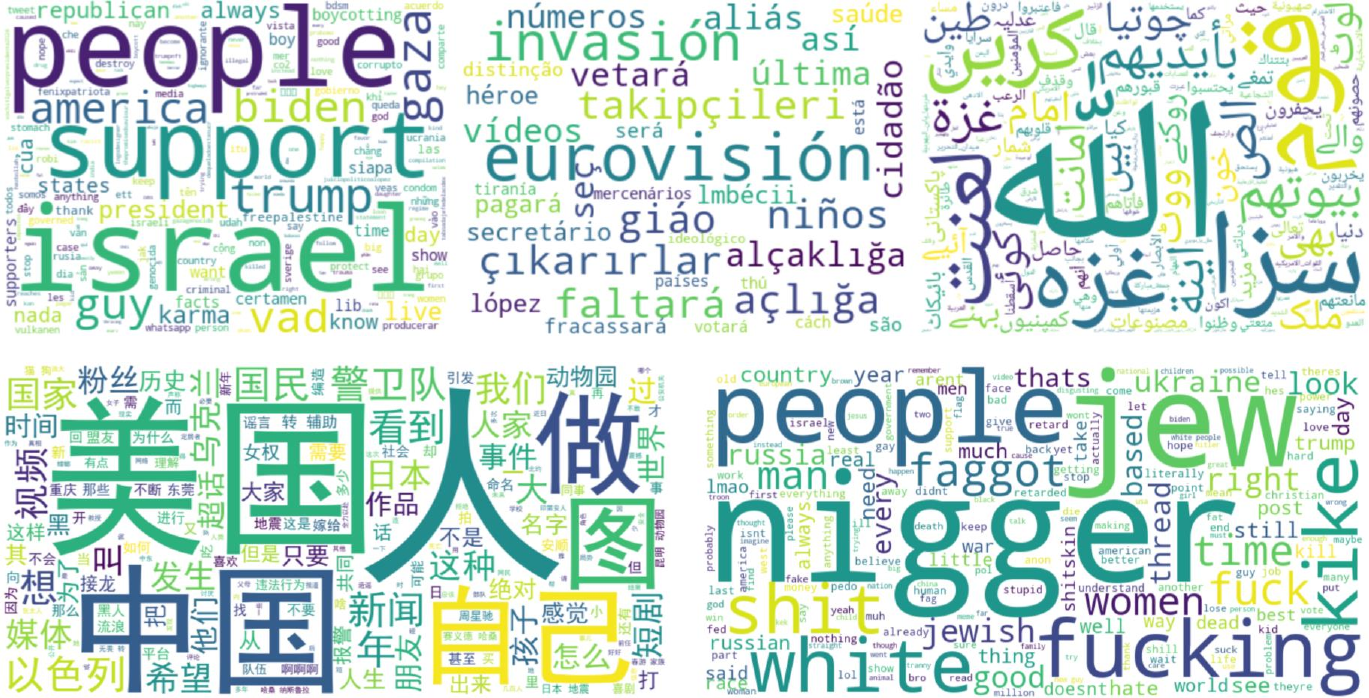} 
\caption{Visualizing linguistic patterns in M\textsuperscript{3}. The top panel displays the word cloud of posts in hate samples from X, while the bottom-left and bottom-right panels illustrate the word cloud of Weibo and 4chan, respectively.}
\label{fig:wordcloud}
\end{figure}

\subsection{Overview}

Following rigorous annotation and filtering, M\textsuperscript{3} comprises 2,455 multimodal samples with 1,400 from 4chan, 526 from X, and 529 from Weibo, covering multiple languages. Each sample consists of an image paired with an accompanying textual post. On average, the post length is 125.96 characters, with a maximum of 789 and a minimum of 20. M\textsuperscript{3} is balanced across the top-level labels, comprising 1,318 hate samples and 1,137 normal samples. 

\subsection{Multi-lingual and Multi-Platform Diversity}

\paragraph{Motivation for Platform Selection.}
We select platforms that differ substantially in moderation intensity, linguistic coverage, and cultural style to increase data diversity.

\begin{itemize}
    \item \textbf{X:} As a global social media platform with rapid information diffusion~\cite{ferrara2016rise}, X provides multilingual content spanning English, Arabic, and other Latin scripts, contributing to critical linguistic diversity.
    \item \textbf{Weibo:} As one of the largest social media platforms in China~\cite{li2023characterizing}, Weibo serves as a primary source of large-scale Chinese-language multimodal content, thereby enriching the cultural diversity.
    \item \textbf{4chan (/pol/):} The /pol/ board on 4chan is characterized by anonymity and minimal moderation, resulting in a high density of extreme and explicit hate content~\cite{colley2022challenges}. It allows M\textsuperscript{3} to capture the upper bound of hateful visual–textual expressions rarely observed on mainstream platforms.
\end{itemize}

M\textsuperscript{3} encompasses a wide variety of languages, such as English, Chinese, and Arabic (details in Section \ref{sec:app-dataset} of the appendix), reflecting the globalized nature of online hate speech. As shown in Figure \ref{fig:wordcloud}, the word cloud reveals pronounced multi-platform heterogeneity in hate expressions, indicating that M\textsuperscript{3} spans a wide spectrum of hate explicitness. On 4chan, high-frequency terms such as ``nigger'', ``kike'', and ``faggot'' co-occur with explicit profanity (e.g., ``fuck'', ``shit''), indicating direct group targeting and overt dehumanization. In contrast, hate expressions on X are largely embedded within political and conflict-related discourse, with frequent references to entities such as ``Israel'', ``Gaza'', ``Trump'', and ``Biden''. On Weibo, high-frequency terms center on national identities (``Chia'', ``America'', ``Japan''), indicating that hate is predomaintly articulated through event-driven discourse.

\begin{figure}[t]
\centering
\includegraphics[width=0.95\linewidth]{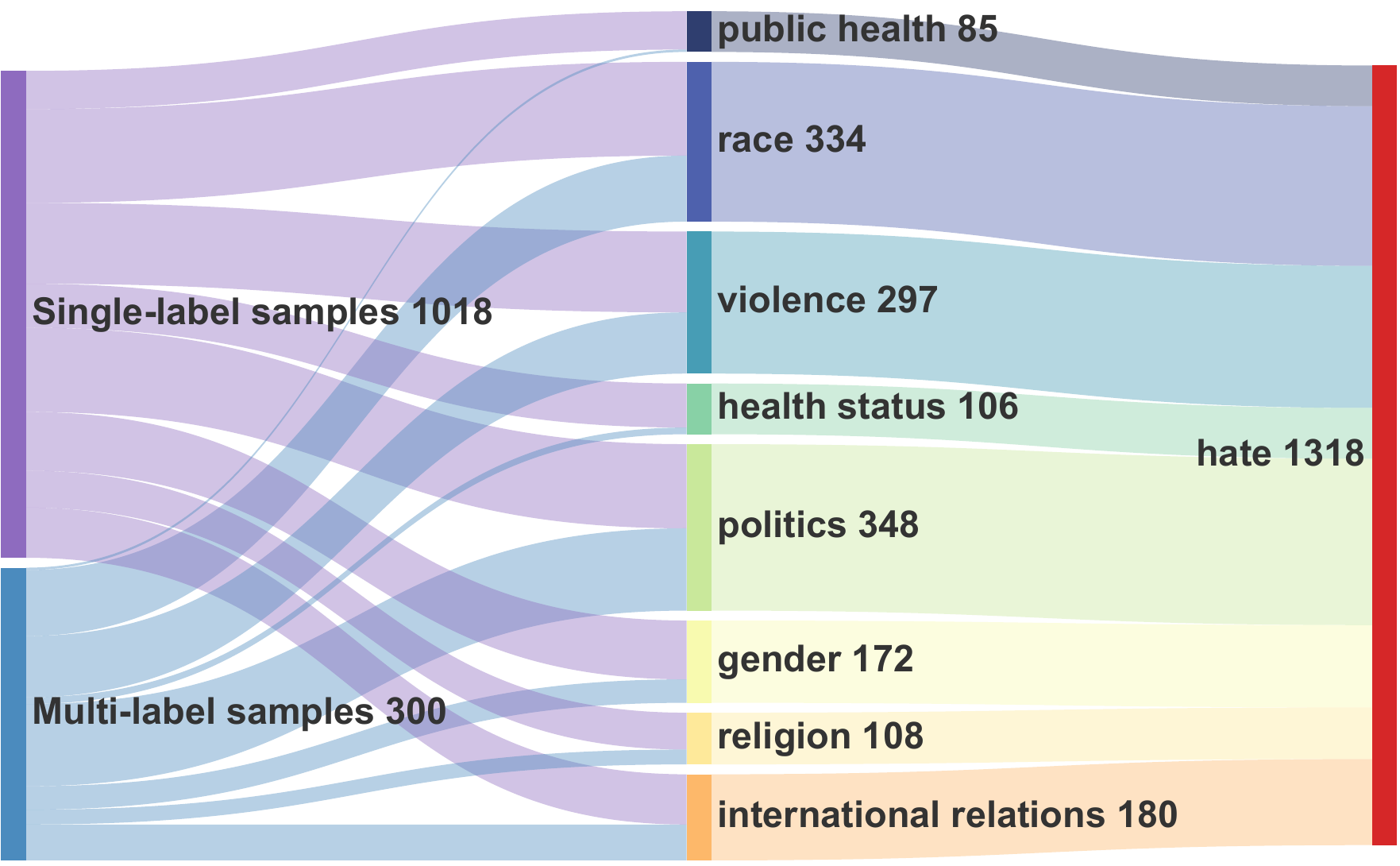}
\caption{Hierarchical categories in M\textsuperscript{3}. Hate samples are categorized into eight themes, with 1,018 single-labeled and 300 multi-labeled samples.}
\label{fig:dataset}
\end{figure}

\subsection{Hierarchical Multi-label Categorization}

\paragraph{Category Definition.} We further operationalize the definition of hate speech by UN~\cite{un_hate_speech_2019} into eight thematic categories to facilitate fine-grained analysis. Each category represents a distinct and socially significant form of hate commonly observed online:

\begin{itemize}
\item \textbf{Religion:} Memes that promote harmful content related to religious conflict, such as disputes or hostility between religious sects or groups.
\item \textbf{Politics:} Memes about political conflict, including hate stemming from government policy failures or partisan disputes, especially during elections.
\item \textbf{Race:} Memes containing racially discriminatory content, targeting individuals or groups based on ethnicity or race.
\item \textbf{Gender: }Memes that involve gender-based discrimination, including harmful content targeting women or the LGBTQ+ community.
\item \textbf{Health Status:} Memes that mock, insult, or discriminate against individuals based on their health conditions, such as disabilities or chronic illnesses.
\item \textbf{Violence:} Memes that incite or glorify acts of violence, encompassing content that explicitly or implicitly advocates targeted shootings, promotes or incites online harassment or cyber violence.
\item \textbf{Public Health:} Memes spreading misinformation or fear-mongering about public health crises, like famine panic or pandemic-related hate speech.
\item \textbf{International Relations:} Memes targeting international relations, for example, provoking inter-country hostility during conflicts or fueling anti-refugee sentiment.
\end{itemize} 

The per-category sample distribution is shown in Figure \ref{fig:dataset}. Notably, due to data collection from the /pol/ board on 4chan, the politics category accounts for $26.4\%$ of hate samples. Among hate samples in M\textsuperscript{3}, $22.76\%$ contain multiple category labels. Statistical analysis shows that the most frequent multi-label combinations include (``race'', ``violence'') and (``international relations'', ``politics'') (58 samples each), followed by (``politics'', race'') and (``politics'', violence'') (40 samples each), which is consistent with intuitive discourse patterns on these platforms.

\begin{figure}[h]
\centering
\includegraphics[width=0.9\linewidth]{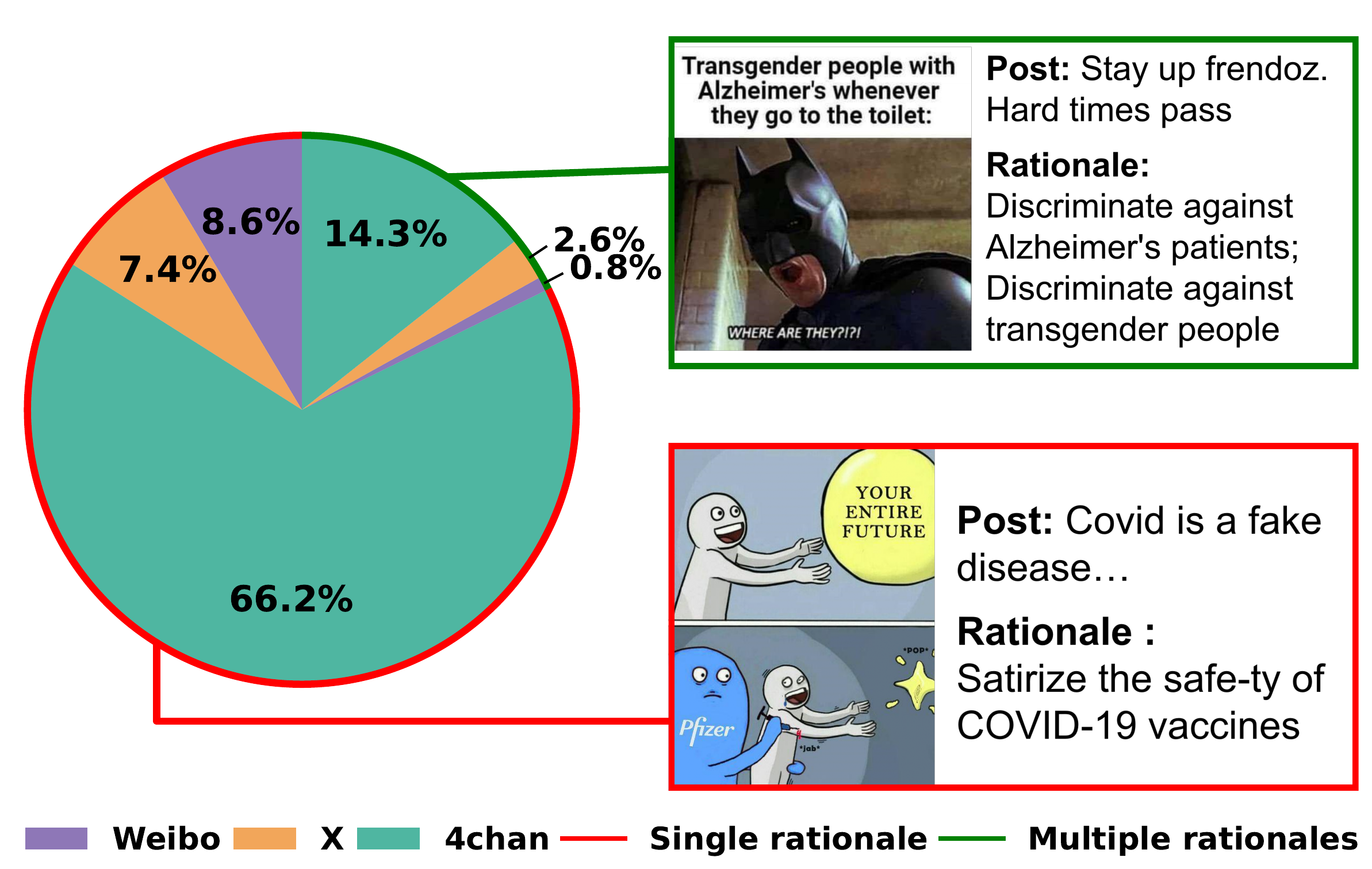} 
\caption{The distribution of single rationale and multiple rationales across X, Weibo, and 4chan.}
\label{fig:reason-distribution}
\end{figure}

\begin{table*}[t]
\renewcommand{\arraystretch}{1.2}
\resizebox{\textwidth}{!}{
\begin{tabular}{cccccccccccccccc}
\toprule
\multirow{2.5}{*}{\textbf{Model}} & \multirow{2.5}{*}{\textbf{Meme}} & \multirow{2.5}{*}{\textbf{Post}} & \multirow{2.5}{*}{Overall$\uparrow$} & \multicolumn{4}{c}{\textbf{Binary}} & \multicolumn{5}{c}{\textbf{Multi-class}} & \multicolumn{3}{c}{\textbf{Rationale}} \\ \cmidrule(lr){5-8} \cmidrule(lr){9-13} \cmidrule(lr){14-16}
 & & & & Acc$\uparrow$ & P$\uparrow$ & R$\uparrow$ & F1$\uparrow$ & Macro-P$\uparrow$ & Macro-R$\uparrow$ & Macro-F1$\uparrow$ & HL$\downarrow$ & Subset Acc$\uparrow$ & BLEU$\uparrow$ & ROUGE$\uparrow$ & BERTScore$\uparrow$ \\ \midrule
\multirow{2}{*}{LLaVA-v1.6-Vicuna-7B-hf} & \usym{2713} & \usym{2717} & 19.83 & 53.69 & 53.69 & \textbf{100.00} & 69.86 & 47.92 & 29.48 & 32.89 & 16.18 & 9.48 & 0.43 & 4.33 & 97.39 \\
 & \usym{2713} & \usym{2713} & 21.82 & 53.69 & 53.69 & \textbf{\underline{100.00}} & 69.86 & 26.40 & 69.46 & 35.61 & 35.07 & 13.20 & 0.06 & 1.89 & 97.41 \\
\multirow{2}{*}{LLaVA-v1.6-Vicuna-13B-hf} & \usym{2713} & \usym{2717} & 19.97 & 66.52 & 61.89 & 97.95 & 75.85 & 47.92 & 29.48 & 32.89 & 16.18 & 9.48 & 0.40 & 5.03 & 96.38 \\ 
 & \usym{2713} & \usym{2713} & 20.03 & 59.47 & 68.84 & 44.76 & 54.25 & 40.75 & 61.39 & 44.75 & 21.48 & 14.26 & 0.29 & 4.51 & 95.90 \\ 
\multirow{2}{*}{GLM-4.1V-9B-Thinking} & \usym{2713} & \usym{2717} & 77.42 & 76.74 & \textbf{98.19} & 57.74 & 72.72 & 58.70 & \textbf{86.50} & \textbf{69.02} & \textbf{11.66} & 40.52 & 1.04 & 8.67 & \textbf{97.76} \\
 & \usym{2713} & \usym{2713} & 80.61 & 77.11 & \textbf{\underline{92.09}} & 62.75 & 74.64 & 54.68 & 81.98 & \textbf{\underline{64.49}} & \textbf{\underline{13.96}} & \textbf{\underline{33.76}} & 1.19 & 9.17 & \textbf{\underline{97.79}} \\ 
\multirow{2}{*}{Qwen2.5-VL-3B-Instruct} & \usym{2713} & \usym{2717} & 51.63 & 76.21 & 78.63 & 76.48 & 77.54 & 56.51 & 75.56 & 62.00 & 14.98 & 28.91 & 0.33 & 3.71 & 97.19 \\
 & \usym{2713} & \usym{2713} & 58.41 & 74.75 & 88.61 & 60.77 & 72.10 & 51.84 & 74.47 & 58.33 & 16.27 & 24.13 & 0.25 & 5.24 & 97.07 \\
\multirow{2}{*}{Qwen2.5-VL-7B-Instruct} & \usym{2713} & \usym{2717} & 82.19 & \textbf{90.26} & 91.15 & 90.67 & \textbf{90.99} & 59.08 & 79.50 & 65.54 & 12.42 & 39.53 & 0.85 & 7.19 & 97.68 \\
 & \usym{2713} & \usym{2713} & 77.43 & \textbf{\underline{86.48}} & 86.63 & 88.47 & \textbf{\underline{87.54}} & \textbf{\underline{56.48}} & 73.64 & 59.59 & 14.74 & 30.80 & 0.64 & 6.63 & 97.34 \\
\multirow{2}{*}{Qwen3-VL-8B-Instruct} & \usym{2713} & \usym{2717} & \textbf{87.63} & 86.80 & 91.76 & 82.85 & 87.08 & \textbf{61.71} & 79.86 & 68.17 & 11.86 & \textbf{42.11} & 1.48 & 9.18 & 97.51 \\
 & \usym{2713} & \usym{2713} & \textbf{\underline{84.83}} & 85.95 & 87.05 & 86.72 & 86.89 & 55.36 & 76.64 & 62.53 & 14.63 & 33.16 & 1.50 & 8.64 & 97.42 \\
\multirow{2}{*}{GPT-4o} & \usym{2713} & \usym{2717} & 60.37 & 86.27 & 94.17 & 80.08 & 86.56 & 50.6 & 79.87 & 60.93 & 15.16 & 26.26 & 0.10 & 7.95 & 97.47 \\
 & \usym{2713} & \usym{2713} & 62.96 & 85.47 & 83.65 & 89.46 & 86.45 & 49.04 & 62.57 & 53.40 & 16.04 & 19.89 & 0.14 & 7.87 & 97.43 \\ 
\multirow{2}{*}{Gemini-3} & \usym{2713} & \usym{2717} & 59.42 & 66.97 & 92.75 & 41.73 & 57.56 & 53.97 & 85.63 & 64.60 & 15.24 & 33.23 & \textbf{1.74} & \textbf{11.64} & 97.28 \\
 & \usym{2713} & \usym{2713} & 75.16 & 73.44 & 86.43 & 59.94 & 70.79 & 48.18 & \textbf{\underline{84.42}} & 59.74 & 18.20 & 25.17 & \textbf{\underline{2.17}} & \textbf{\underline{13.07}} & 97.51 \\
\bottomrule
\end{tabular}
}
\caption{Comparison of state-of-the-art MLLMs on the M\textsuperscript{3} dataset. The models are listed in order from open-source to proprietary, following their chronological release or version evolution: (1) the early LLaVA-v1.6 series (7B, 13B), (2) the reasoning-enhanced GLM-4.1V-9B-Thinking, (3) the latest Qwen series (Qwen2.5-VL-3B/7B and Qwen3-VL-8B), and (4) the proprietary GPT-4o and Gemini-3. Results cover binary and multi-class classification, and rationale quality under meme-only and meme + post settings. Boldface denotes the extremal values under the meme-only setting, while boldface with underline denotes the extremal values under the meme + post setting. }
\label{tab:performance}
\end{table*}

\subsection{Rationales for hate memes}

For the hate samples, we annotate 1,557 rationales describing why the content is hateful. Each rationale follows a \textit{$<$verb$>$$<$object$>$} structure (e.g., insult black people), with an average length of 32.35 characters, the longest being 101, and the shortest 9. As shown in Figure \ref{fig:reason-distribution}, hate samples are annotated with a single rationale, while samples associated with multiple rationales constitute a smaller portion across all platforms. The three most common rationales include ``express political hatred'' (44 times in the politics category), ``depreciate transgender individuals'' (33 times in the gender category), and ``discriminate against people with intellectual disabilities'' (29 times in the health status category). 

In summary, M\textsuperscript{3} is a thematically diverse multimodal benchmark with broad coverage. Its inclusion of real-world social media contexts, hierarchical multi-label annotations, and comprehensive rationales makes it a valuable resource for evaluating the nuanced hate recognition and interpretation capabilities of MLLMs, especially within the real-world dynamic environments.

\section{Experiments}

\subsection{Experiment Setups}

\paragraph{Dataset and Baselines.} Experiments are conducted on M\textsuperscript{3}, containing 2,455 memes paired with corresponding posts. Each instance is annotated with binary labels, and instances labeled as hate are further annotated with fine-grained categories and phrase-level rationales. We compare several representative MLLMs, categorized into open-source and proprietary models, and ordered by their release dates or versions: (1) Open-source models include LLaVA-v1.6 series (7B and 13B, Feb 2024), GLM-4.1V-9B-Thinking (Jul 2025), Qwen2.5-VL series (3B and 7B, Jan 2025), Qwen3-VL-8B-Instruct (Oct 2025); (2) Proprietary models include GPT-4o (May 2024) and Gemini-3 (Nov 2025). GPT-4o and Gemini-3 are accessed via the official API, whereas the other models are deployed locally.

\paragraph{Tasks and Evaluation Metrics.} We investigate three meme understanding tasks: (1) Binary hate detection, which predicts whether a meme contains hateful content (hate vs. normal); (2) Multi-label fine-grained classification, identifying specific categories present in the hateful meme; and (3) Rationale generation, which produces a concise verb–object phrase explaining why the meme is hateful. For binary classification, we report Accuracy, Precision, Recall, and F1-score. For multi-label fine-grained classification, we compute macro-Precision, macro-Recall, macro-F1, Hamming Loss, and Subset Accuracy~\cite{zhang2013review}. For rationale generation, we assess the quality of generated rationales using BLEU~\cite{papineni2002bleu}, ROUGE~\cite{lin2004rouge} and BERTScore~\cite{zhang2019bertscore}. We define an Overall score by first normalizing all metrics to $[0,1]$ (inverting where necessary) and taking average across the three tasks. 

\paragraph{Implementation Details.} For each model, we explore two input settings: (i) Meme-only, where only the meme image is provided; and (ii) Meme + Post, where the meme is paired with its associated post text. All experiments are conducted on a single NVIDIA A800 GPU (80GB). To ensure fair comparison, no task-specific fine-tuning is performed; instead, models are directly evaluated on downstream tasks in a zero-shot setting.

\subsection{Results and Discussion}

\begin{figure*}[t]
\centering
\includegraphics[width=0.95\textwidth]{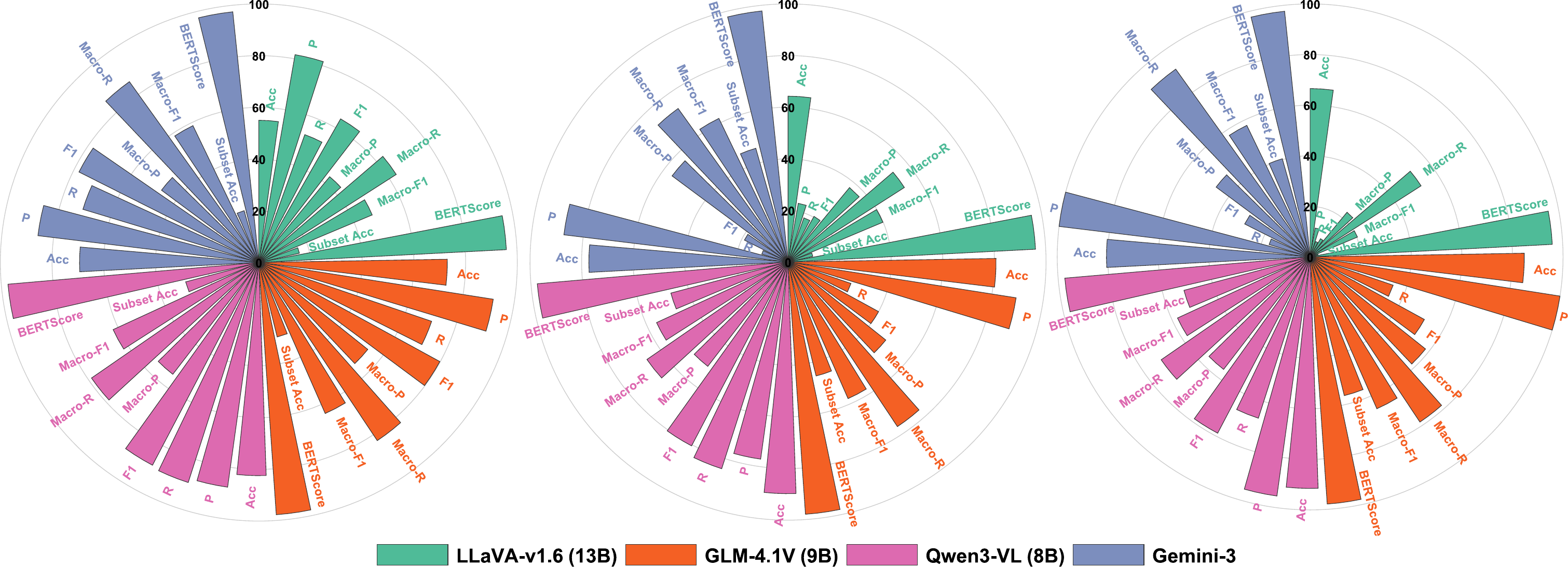}
\caption{Performance comparison of MLLMs across three platforms. From left to right: 4chan (English-centric), X (Multi-lingual, e.g., Latin and Arabic), and Weibo (Chinese-dominant).}
\label{fig:platform-results}
\end{figure*}

\paragraph{Different Tasks.} According to Table \ref{tab:performance}, a clear task-level performance hierarchy emerges across all evaluated models. models achieve robust results in binary classification (typically $>85\%$ accuracy) but struggle with multi-class tasks (Macro-F1: $32.89\%$–$69.02\%$). Notably, scaling model capacity yields non-uniform gains. For example, increasing model capacity from LLaVA-7B to LLaVA-13B yields only marginal improvements in binary accuracy but leads to a noticeable increase in multi-class macro-F1. Similarly, within the Qwen family, scaling from 3B to 7B results in clear gains in binary classification (+13.7 accuracy points), whereas further scaling to Qwen3-VL-8B produces diminishing returns for binary accuracy but more pronounced benefits for multi-class macro-F1 and rationale metrics. A unique pattern emerges in rationale generation where high BERTScore values (above 95.0) coexist with low lexical-overlap metrics, including BLEU scores below 2.2 and a peak ROUGE of 11.64 (Gemini-3, meme-only). This discrepancy arises because BLEU and ROUGE operate at the word level, failing to capture the semantic alignment of our generated rationales, which primarily consist of verb-object phrases with an average length of 32.35 characters.

\paragraph{Multimodal Inputs.} As shown in Table \ref{tab:performance}, incorporating posts alongside memes produces heterogeneous effects across models and tasks, yielding inconsistent performance gains across the evaluated dimensions. Adding post information yields a negligible impact on binary classification (F1 $\pm 3$ points) but triggers a consistent performance decline in multi-class tasks. Specifically, macro-F1 scores drop in 6 of 9 evaluated models, with significant in high-performing models like Qwen3-VL and GPT-4o. Although LLaVA showed an 11.86-point improvement, this behavior represents an exception rather than the dominant pattern. Across newly released models, BLEU and ROUGE improve when post text is included, suggesting that posts provide complementary semantic cues that facilitate rationale construction. Taken together, these results indicate that current MLLMs struggle to integrate meme and post information for fine-grained intent understanding robustly. While additional context may help rationale generation, it often introduces ambiguity that degrades classification performance, highlighting a key limitation in real-world hate speech moderation scenarios where user intent is distributed across modalities.

\paragraph{Multi-platform and Multi-lingual Evaluation.} We select four multimodal models (LLaVA-v1.6-Vicuna-13B-hf, GLM-4.1V-9B-Thinking, Qwen3-VL-8B-Instruct, and Gemini-3) from different model families that demonstrate strong overall performance in prior experiments and evaluate them across 4chan, X, and Weibo. A pronounced precision-recall trade-off emerges on X and Weibo, particularly for Gemini-3, which achieves perfect precision but a meager 16.94 recall on Weibo. These results suggest that certain models adopt overly conservative prediction strategies, prioritizing precision while sacrificing recall. Although this reduces false positives, it leads to a large number of false negatives and limits practical usefulness. The effect is particularly pronounced under domain and language shifts, indicating that alignment or thresholding mechanisms may bias models toward excessively ``safe'' predictions in multi-platform settings. From a multi-lingual perspective, models generally perform better on English-dominated platforms (4chan) compared to non-English or mixed-language platforms like X and Weibo. Performance degradation is particularly noticeable for Chinese-language content, where high-precision models like Gemini-3 still suffer from dramatic recall drops, highlighting challenges in cross-linguistic generalization. For clarity, we visualize only the results under the meme + post setting. Results are reported in full in Section \ref{sec:app-multi} of the appendix.

\section{Conclusion}

In this work, we introduce \textbf{M\textsuperscript{3}}, a multi-platform, multi-lingual, and multimodal meme dataset constructed through an agentic annotation framework with systematic human verification. M\textsuperscript{3} contains 2,455 multimodal instances from X, 4chan, and Weibo, annotated with binary hate labels, fine-grained categories, and human-verified rationales. We benchmark M\textsuperscript{3} on state-of-the-art MLLMs. Results show that incorporating surrounding post context does not consistently improve hate detection and often degrades fine-grained classification, although it can benefit rationale generation. These findings highlight current limitations of MLLMs in integrating multimodal contextual information. Overall, M\textsuperscript{3} provides a challenging benchmark for multimodal hate speech analysis and offers insights into the gap between existing model capabilities and real-world moderation needs.

\appendix

\section*{Ethical Statement}

This work involves the analysis of hateful memes that may contain offensive content. All data in M\textsuperscript{3} are collected from publicly available platforms and are used solely for research purposes. Personally identifiable information is removed during data processing.

Annotations are conducted with human verification, and annotators are informed of the sensitive nature of the content. This study does not endorse hateful expressions. The dataset is intended to support research on multimodal hate detection and should be used responsibly in accordance with ethical guidelines.

\section*{Acknowledgments and Disclosure of Funding}
This work was supported in part by the National Natural Science Foundation of China (62306229), the Youth
Talent Support Program of Shaanxi Science and Technology Association (20240113), the China
Postdoctoral Science Foundation (2025T180425).

\bibliographystyle{named}
\bibliography{ijcai26}

\clearpage
\appendix
\section{The Details of the Agentic Annotation Framework}

This section provides detailed implementation and reproducibility information for the Agentic Annotation Framework (Section \ref{sec:framework}), organized according to the four main processes used to construct the M\textsuperscript{3} dataset: Data Acquisition, Preprocessing, Multimodal Annotation, and Expert-driven Quality Assurance.

\subsection{Process 1: Data Acquisition}

\paragraph{Collector.} 
The Collector harvests raw images and associated JSON metadata (Table~\ref{tab:json_fields}) from three diverse and complementary social platforms: X (formerly Twitter), Weibo, and 4chan’s /pol/ board. This selection ensures a broad linguistic and cultural scope, capturing a wide spectrum of image-based hate speech. We collected 43,567 image-post pairs including English, Arabic, Japanese, and other Latin and non-Latin script languages from X. The smaller volume is due to limited API stability and access constraints during our specific collection window in February 2024. 

\begin{table}[ht]
\centering
\resizebox{\linewidth}{!}{
    \begin{tabular}{ll}
    \toprule
    \textbf{Field} & \textbf{Description} \\ \midrule
    \texttt{post\_id}      & Unique identifier of the post \\
    \texttt{post\_time}    & Timestamp the post was published \\
    \texttt{img\_id}       & Unique identifier for the meme \\
    \texttt{img\_url}      & URL linking to the meme \\
    \texttt{post\_text}    & Cleaned textual content of the post \\
    \texttt{img\_text}     & OCR-extracted textual content within the meme \\  \bottomrule
    \end{tabular}
    }
    \caption{Fields in JSON Metadata Files. \texttt{post\_text} contains the textual content of the post after Extractor and Cleaner.}
    \label{tab:json_fields}
\end{table}

\subsection{Process 2: Preprocessing}

\paragraph{Extractor.}  
The Extractor uses PaddleOCR to extract embedded image text (\texttt{img\_text}) and combines it with post text (\texttt{post\_text}) to create unified textual representations. Processing is parallelized, skips images with pre-existing OCR results, and aggregates line-level outputs into single strings per image:

\begin{itemize}
    \item \textbf{Batch Processing and Parallelism:} Images are processed in parallel using a thread pool executor, which enables concurrent OCR inference to accelerate throughput on multi-core systems.
    \item \textbf{OCR Model Configuration:} PaddleOCR is initialized with angle classification enabled and set to recognize English text, allowing detection of rotated or stylized text commonly found in memes.
    \item \textbf{Incremental Processing:} For each image, the corresponding JSON metadata file is checked for existing OCR results. If recognized text is already present, the image is skipped to avoid redundant computation.
    \item \textbf{Text Extraction and Aggregation:} The OCR outputs are parsed to extract line-level recognized text, which are concatenated with newline separators to form a single textual string representing the image content.
    \item \textbf{Result Persistence and Error Logging:} The extracted text is appended to the image's JSON metadata under the \texttt{recognized\_text} field. Any OCR processing failures are logged into a dedicated failure record file for later inspection and potential reprocessing.
\end{itemize}

\paragraph{Cleaner.}  
To reduce noise and enhance the semantic quality of textual data, we apply platform-specific cleaning procedures to both the post content and the OCR-extracted embedded text. These include:

\begin{itemize}
    \item Removal of URLs and web links, which are common but irrelevant for hate speech semantics.
    \item Elimination of user mentions (e.g., \texttt{@username}), with distinct regular expressions adapted to each platform's syntax to accurately remove references without harming context.
    \item Filtering out hashtags, including specially formatted tags on platforms like Weibo (e.g., \texttt{\#...\#} or \texttt{[\#...\#]}), to focus on natural language content.
    \item Replacement of newlines and excessive whitespace with standardized delimiters (commas or spaces) to normalize textual structure and facilitate downstream processing.
    \item For 4chan data, additional removal of quoting symbols and post references (e.g., \texttt{>>12345}) that do not contribute to meme semantics.
\end{itemize}

These tailored cleaning steps ensure that the textual modalities reflect core semantic information relevant for hate speech detection, while minimizing platform-specific noise artifacts.

Given the heterogeneous nature of user-generated content, we impose length-based thresholds on the combined textual representation (post content plus embedded image text) to retain only samples likely to carry meaningful semantic content. Specifically:

\begin{itemize}
    \item Samples with overly short text (e.g., fewer than 20 characters for 4chan posts, or fewer than 10 characters for embedded image text) are discarded as they lack sufficient context for interpretation.
    \item Samples with excessively long text (e.g., beyond 500 characters for 4chan post content, or 100 characters for embedded text) are also excluded to avoid noise from spam, off-topic content, or multi-topic posts.
    \item Corresponding image files are removed in tandem with filtered JSON metadata to maintain dataset integrity.
\end{itemize}

This length-based filtering step balances data quality and scale, ensuring the final dataset consists of content-rich multimodal memes suitable for detailed annotation and model evaluation.

Following this pipeline, we retain 16,179 meme samples from X, 298,043 from Weibo, and 221,249 from 4chan.

\subsection{Process 3: Multimodal Annotation}
\label{sec:app-annotation}
\paragraph{Annotator.}  
Three MLLMs (GLM-4v-flash, llava-v1.6-mistral-7b-hf, Qwen2.5-VL-7B-Instruct) as the Annotator predict for each meme, including a binary hate label (Normal, Hate) and Fine-grained category label(s) (Religion, Politics, Race, Gender, Health Status, Violence, Public Health, International Relations). Specifically, the three models come from different model families and training paradigms, providing complementary strengths in multimodal reasoning, instruction following, and cross-lingual understanding: GLM-4v-flash is optimized for fast and low-cost inference, LLaVA-v1.6-Mistral-7B represents a strong open-source baseline with competitive multimodal understanding, and Qwen2.5-VL-7B-Instruct offers robust vision–language alignment and multilingual performance. Restricting the ensemble to three models significantly reduces inference cost and latency, which is crucial for large-scale meme annotation

We adopt a prompt-based annotation strategy to guide the annotators. The specific prompts used for annotation are as follows:

\begin{promptbox}
\begin{description}
    \item[\textbf{P1}] You are a hateful meme detection assistant.
    Based on the input image and post: \{post\_text\}, please select the most appropriate option from [``hate", ``normal"].
    
    And give your confidence score for this option as a decimal between 0 and 1.
    
    Please output in this format:\{``label": ``hate" or ``normal", ``confidence\_score": 0.0-1.0\}

    \item[\textbf{P2}] You are a harmful meme classification assistant.
    Based on the the content of the image, please select the most appropriate one or more of the following categories: 
    
    Category list: [``religion", ``politics", ``race", ``gender", ``health status", ``violence", ``public health", ``international relations"].
    
    Please output in json format: \{"category": [``category1", ``category2"], ``confidence\_score": [0.78, 0.56]\}
\end{description}
\end{promptbox}

\paragraph{Arbiter.} 
\begin{figure*}[t]
\centering
\includegraphics[width=0.95\textwidth]{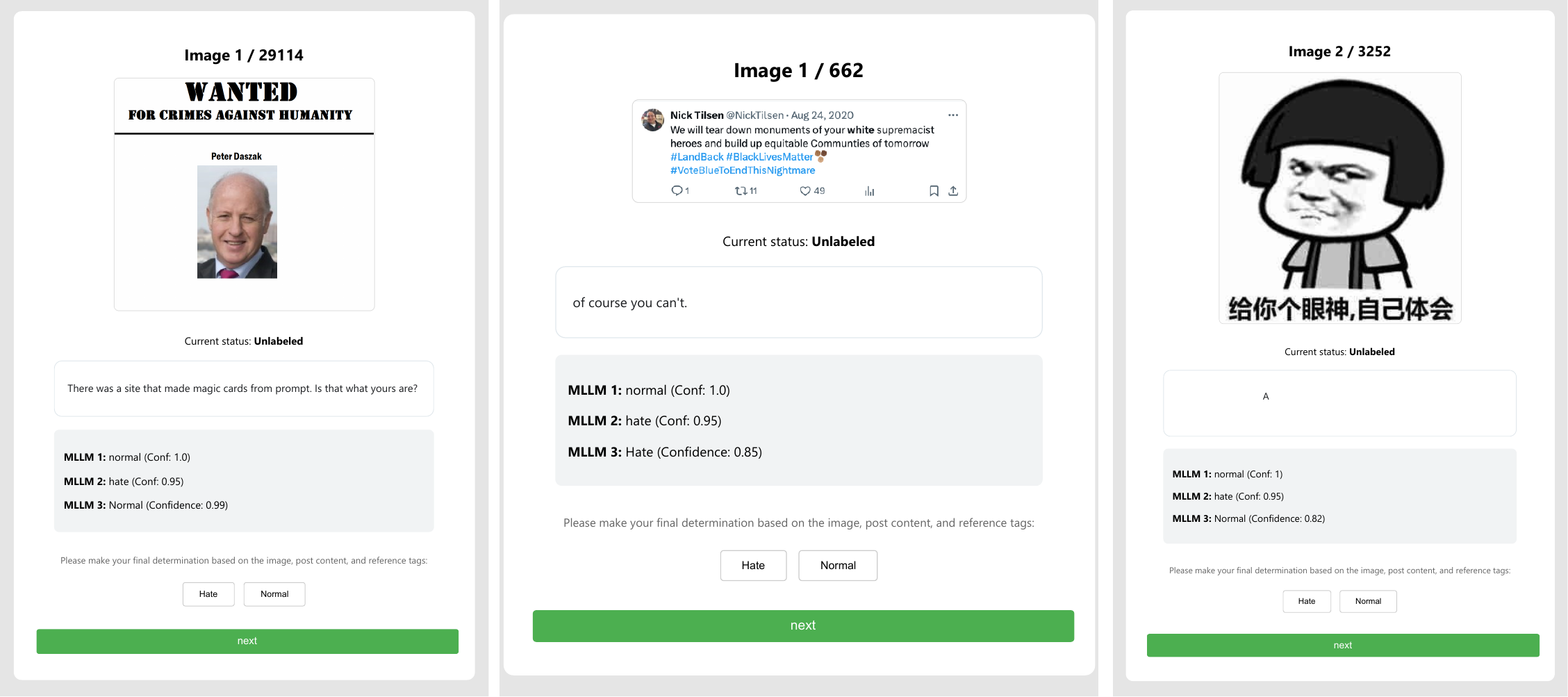}
\caption{Web-based arbitration interface for binary hate annotation. From left to right, examples are sourced from 4chan, X, and Weibo. The interface displays the meme alongside predictions from annotators to support human review of low-consistency samples.}
\label{fig:hate-annotation}
\end{figure*}

\begin{figure*}[t]
\centering
\includegraphics[width=0.95\textwidth]{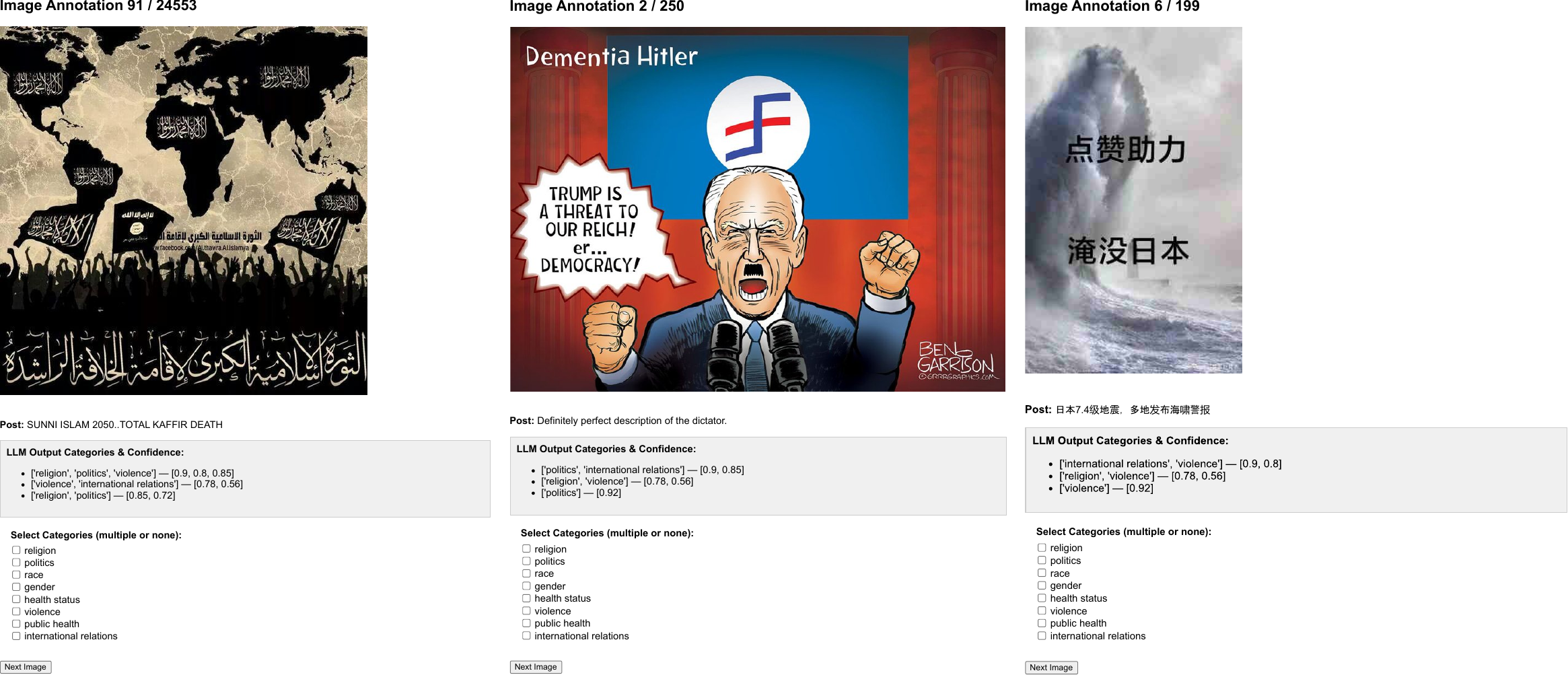}
\caption{Web-based arbitration interface for hate category annotation. From left to right, examples are sourced from 4chan, X, and Weibo. The interface presents hate categories with suggestions of annotators, allowing human arbiters to select, combine, or override model predictions.}
\label{fig:category-annotation}
\end{figure*}
The web-based interface is used only for manual review of low-consistency samples identified by the Arbiter. It displays the meme image together with predictions from multiple LLM annotators for two tasks: binary hate labeling (see Figure \ref{fig:hate-annotation}) and hate category annotation (see Figure \ref{fig:category-annotation}). As a representative example, the category arbitration interface (Figure \ref{fig:hate-annotation}) presents eight predefined hate categories. For each meme, the predictions from three LLMs are displayed as checkbox options, allowing human arbiters to select one or more suggestions, combine them, or reject all and leave the label blank. Upon submission, the results are written incrementally to a local JSON file (\texttt{final\_annotations.json}), and the interface automatically proceeds to the next sample.

This interface is built on the Flask web framework and utilizes dynamic HTML templates for rendering. Static image assets are served locally to ensure low latency and data security during annotation. Hate label selection and category correction modules follow the same annotation flow and backend architecture, differing only in label type and instruction text.

\subsection{Process 4: Expert-driven Quality Assurance}

\paragraph{Explicator.} 
The Explicator generates structured verb-object rationales for hate memes by combining visual and textual cues. First, a prompt is given to GPT-4o to produce an initial rationale or output. The prompt instructs the model to strictly output rationales for a meme’s harmfulness using only verb-object phrases, separated by commas, with a maximum of 30 words. 

\begin{promptbox}
    You are a hateful meme explicator. \\
    Strictly output rationale for hate in English, using only verb-object phrases. \\
    If the meme have many rationales, separate verb-object phrases with commas. Do not add anything else. No more than 30 words. \\
    Given: \\
    - post:\{post\_content\} \\
    - category:\{category\}
\end{promptbox}

The output of the model serves as a draft, which may contain imprecise wording, incomplete logic, or formatting inconsistencies. Humans then review and edit this draft to correct verb choices, clarify objects, ensure word limits, and enforce the specified format. During manual refinement stage, explicators are are provided with explicit guidance, including the attack target (individual, group, society, and nation) and the attack type (abuse, discriminate, ridicule, dehumanize, and incite violence), to ensure accurate editing.

\paragraph{Validator.}
To further assess the reliability of high-consistency samples, we conducted a random audit of 200 instances that had unanimous agreement in the arbitration. Table \ref{tab:votes} reports the detailed voting outcomes of three graduate validators. Each validator independently marked agreement (1) or opposition (0) for every sample. At the individual level, each validator maintained high agreement rates, ranging from 94.5\% to 97.5\%, with opposition limited to 5–11 cases per validator. These results confirm the robustness of the Arbiter’s initial decisions. 

\begin{table}[t]
\centering
\resizebox{\linewidth}{!}{
\begin{tabular}{lcc}
\toprule
\textbf{Validator} & \textbf{Agreement (1)} & \textbf{Opposition (0)} \\ 
\midrule
Graduate 1         & 195                     & 5                       \\
Graduate 2         & 194                     & 6                       \\
Graduate 3         & 189                     & 11                      \\ 
\midrule
\multicolumn{3}{c}{\textbf{Vote Statistics}} \\ 
\midrule
Three votes in favor & \multicolumn{2}{c}{181} \\ 
Two votes in favor   & \multicolumn{2}{c}{16}  \\ 
One vote in favor    & \multicolumn{2}{c}{3}   \\ 
No votes in favor    & \multicolumn{2}{c}{0}   \\ 
\bottomrule
\end{tabular}
}
\caption{Validator voting results for a random audit of 200 high-consistency samples. The table reports individual agreement/opposition counts and aggregated vote statistics.}
\label{tab:votes}
\end{table}

\section{Dataset Details and Examples}
\label{sec:app-dataset}

In this section, we provide a supplementary overview and analysis of our dataset M\textsuperscript{3}. We describe its overall composition, representative examples and detailed multi-lingual statistics.

\subsection{Representative Examples per Category}

Table \ref{tab:single_label_examples} presents representative examples across the eight hate-related categories. Each entry includes a meme identifier, the post text, the annotated category, and a rationale explaining its hateful content.

These examples highlight the spectrum of hateful expression in social and political contexts, demonstrating that effective detection requires identifying not only toxicity but also the specific thematic target. M\textsuperscript{3} captures a range of rhetorical strategies, from overt slurs and explicit hostility (\texttt{ID 714}, \texttt{ID 653}) to satire, indirect blame, and ideological dog whistles (\texttt{ID 405}, \texttt{ID 1104}), necessitating nuanced cultural and contextual understanding. Consequently, robust detection must integrate multimodal and sociolinguistic inference.

Category-level distinctions further underscore varied manifestations of hate: attacks on marginalized identities (e.g., race, gender, health status) often involve dehumanization or invalidation, whereas political and international content frequently conveys ideological vilification or geopolitical mockery.

\begin{table*}[t]
\centering
\renewcommand{\arraystretch}{1.2}
\resizebox{\textwidth}{!}{
\begin{tabular}{c c p{8cm} p{6cm}}
\toprule
\textbf{ID} & \textbf{Category} & \textbf{Post} & \textbf{Rationale} \\
\midrule
405 & Religion & Everyone but Christians refuse to participate in usury. & Blame Christians while using apocalyptic language \\
502 & Politics & Mana. I had such high hopes for her... Then she came out as a commie. & Incite anti-communist hatred \\
653 & Race & If there's one speck of... is it still just good ol'e milk? & Insult Black people \\
714 & Gender & Trannies be like ``it’s not a mental illness''... & Depreciate transgender individuals \\
832 & Health Status & A big thing I used to do... but just for me these days. & Discriminate against people with depression \\
930 & Violence & Can't come soon enough & Encourage suicide \\
1011 & Public Health & We care... watch you cope in your 55K a year shill cubicle. & Spread vaccine death conspiracy \\
1104 & International Relations & No freedom bucks for you & Satirize Europe's dependence on the United States \\
\bottomrule
\end{tabular}
}
\caption{Representative single-label examples for each hate category.}
\label{tab:single_label_examples}
\end{table*}

\subsection{Multi-lingual Diversity}

\begin{table}[h]
\centering
\resizebox{\linewidth}{!}{
    \begin{tabular}{l r}
    \toprule
    \textbf{Statistics} & \textbf{Value} \\
    \midrule

    \multicolumn{2}{l}{\textbf{Post Text Language Distribution}} \\
    \midrule
    Total posts & 2455 \\
    Monolingual posts & 2117 \\
    \quad * Monolingual -- Latin & 1794 \\
    \quad * Monolingual -- Chinese & 323 \\
    Multilingual posts & 338 \\
    \quad * Multilingual -- Chinese + Latin & 216 \\
    \quad * Multilingual -- Arabic + Latin & 101 \\
    \quad * Multilingual -- Chinese + Japanese + Latin & 15 \\
    \quad * Multilingual -- Korean + Latin & 3 \\
    \quad * Multilingual -- Arabic + Hebrew + Latin & 1 \\
    \quad * Multilingual -- Chinese + Korean + Latin & 1 \\
    \quad * Multilingual -- Chinese + Cyrillic + Latin & 1 \\

    \midrule
    \multicolumn{2}{l}{\textbf{Image Text Language Distribution}} \\
    \midrule
    Total images & 2455 \\
    Monolingual images & 2261 \\
    \quad * Monolingual -- Latin & 1993 \\
    \quad * Monolingual -- Chinese & 268 \\
    Multilingual images & 194 \\
    \quad * Multilingual -- Chinese + Latin & 194 \\
    
    \midrule
    \multicolumn{2}{l}{\textbf{Post-Image Language Comparison}} \\
    \midrule
    Language-aligned pairs & 2326 \\
    Language-misaligned pairs & 129 \\
    Alignment ratio & 94.7\% \\
    Misalignment ratio & 5.3\% \\
    \bottomrule
    \end{tabular}
    }
\caption{Language statistics of post and image text.}
\label{tab:language-statistics}
\end{table}

To examine the linguistic and cultural diversity of the dataset, we analyze post and image text using a Unicode script–based approach. Latin-script languages are grouped together, and posts containing multiple scripts are labeled as multilingual (Table \ref{tab:language-statistics}).

Most post texts are monolingual, dominated by Latin-script content, followed by Chinese, while multilingual posts often mix Chinese–Latin or Arabic–Latin, with rarer combinations involving Japanese, Korean, Hebrew, and Cyrillic scripts. OCR-extracted image text shows similar patterns: 2,237 monolingual and 194 multilingual samples, primarily Latin and Chinese, with multilingual cases largely Chinese–Latin.

Cross-modal analysis reveals that image and post text are mostly linguistically aligned, though some instances exhibit explicit mismatches, reflecting cross-lingual divergence. Overall, the dataset spans Western, East Asian, Middle Eastern, and Slavic scripts, capturing both monolingual and multilingual communication across modalities.

\section{Experiments}

\subsection{Multi-platform, Multi-lingual, and Multimodal Evaluation.}
\label{sec:app-multi}

\begin{table*}[]
\resizebox{\textwidth}{!}{
    \begin{tabular}{@{}ccccccccccc@{}}
    \toprule
    \rowcolor{gray!30} \multicolumn{11}{c}{Meme + Post}                                                                                      \\ \midrule
    Platform               & Model                    & Acc   & P     & R     & F1    & Macro-P & Macro-R & Macro-F1 & Subset Acc & BERTScore \\ \midrule
    \multirow{4}{*}{4chan} & LLaVA-v1.6-Vicuna-13B-hf & 55    & 81.58 & 52.54 & 63.92 & 43.03   & 63.13   & 47.56    & 16.1       & 95.89     \\
                           & GLM-4.1V-9B-Thinking     & 72.93 & 91.8  & 70.62 & 79.83 & 53.92   & 82.73   & 63.79    & 29.66      & 97.76     \\
                           & Qwen3-VL-8B-Instruct     & 82.43 & 87.71 & 89.36 & 88.53 & 54.56   & 78.66   & 62.06    & 29.28      & 97.41     \\
                           & Gemini-3                 & 69.36 & 86.09 & 71.09 & 77.88 & 46.9    & 86.01   & 58.74    & 20.83      & 97.46     \\ \midrule
    \multirow{4}{*}{X}     & LLaVA-v1.6-Vicuna-13B-hf & 64.26 & 23.08 & 18.18 & 20.34 & 37.42   & 53.55   & 39.87    & 9.85       & 96.04     \\
                           & GLM-4.1V-9B-Thinking     & 80.61 & 89.47 & 25.76 & 40    & 48.47   & 76.56   & 57.79    & 45.45      & 97.94     \\
                           & Qwen3-VL-8B-Instruct     & 89.73 & 77.08 & 84.09 & 80.43 & 50.98   & 66.45   & 56.09    & 46.97      & 97.54     \\
                           & Gemini-3                 & 77.19 & 87.5  & 10.61 & 18.92 & 55.97   & 73.16   & 61.76    & 45.8       & 97.8      \\ \midrule
    \multirow{4}{*}{Weibo} & LLaVA-v1.6-Vicuna-13B-hf & 66.54 & 11.59 & 6.45  & 8.29  & 22.9    & 52.1    & 20.3     & 3.23       & 95.87     \\
                           & GLM-4.1V-9B-Thinking     & 84.69 & 100   & 34.68 & 51.5  & 58.36   & 78.38   & 65.54    & 56.45      & 97.89     \\
                           & Qwen3-VL-8B-Instruct     & 91.49 & 95.4  & 66.94 & 78.67 & 56.09   & 71.56   & 57.72    & 51.61      & 97.35     \\
                           & Gemini-3                 & 80.53 & 100   & 16.94 & 28.97 & 46.29   & 91.29   & 57.7     & 40.32      & 97.61     \\ \bottomrule
    \rowcolor{gray!30} \multicolumn{11}{c}{Meme - Only}                                                                                       \\ \midrule
    Platform               & Model                    & Acc   & P     & R     & F1    & Macro-P & Macro-R & Macro-F1 & Subset Acc & BERTScore \\ \midrule
    \multirow{4}{*}{4chan} & LLaVA-v1.6-Vicuna-13B-hf & 78.29 & 78.07 & 99.25 & 87.4  & 48.25   & 30.99   & 34.19    & 10.55      & 96.36     \\
                           & GLM-4.1V-9B-Thinking     & 70.5  & 98.22 & 62.24 & 76.2  & 58.4    & 87.62   & 68.83    & 38.04      & 97.73     \\
                           & Qwen3-VL-8B-Instruct     & 82.93 & 92.64 & 84.18 & 88.21 & 61.33   & 81.18   & 67.9     & 38.98      & 97.51     \\
                           & Gemini-3                 & 58.71 & 93.06 & 49.25 & 64.41 & 52.9    & 87.08   & 63.89    & 29         & 97.22     \\ \midrule
    \multirow{4}{*}{X}     & LLaVA-v1.6-Vicuna-13B-hf & 51.9  & 34.12 & 98.48 & 50.68 & 46.4    & 20.31   & 26.4     & 6.81       & 96.46     \\
                           & GLM-4.1V-9B-Thinking     & 82.32 & 97.56 & 30.3  & 46.24 & 49.72   & 81.49   & 59.6     & 40.15      & 97.9      \\
                           & Qwen3-VL-8B-Instruct     & 90.49 & 83.61 & 77.27 & 80.31 & 54.56   & 70.02   & 59.3     & 50.76      & 97.6      \\
                           & Gemini-3                 & 76.81 & 91.67 & 8.33  & 15.28 & 54.38   & 70.11   & 60.18    & 46.97      & 97.53     \\ \midrule
    \multirow{4}{*}{Weibo} & LLaVA-v1.6-Vicuna-13B-hf & 49.91 & 30.14 & 86.29 & 44.68 & 38.72   & 16.26   & 16.15    & 3.23       & 96.5      \\
                           & GLM-4.1V-9B-Thinking     & 87.71 & 98.36 & 48.39 & 64.86 & 57.42   & 91.76   & 67.46    & 62.1       & 97.91     \\
                           & Qwen3-VL-8B-Instruct     & 93.38 & 93.2  & 77.42 & 84.58 & 54.87   & 77.97   & 61.45    & 59.68      & 97.47     \\
                           & Gemini-3                 & 79.02 & 84.21 & 12.9  & 22.38 & 50.35   & 92.14   & 62.25    & 54.84      & 97.57     \\ \bottomrule 
    \end{tabular}
}
\caption{Multi-platform and multi-lingual evaluation results under two input settings: \textit{Meme + Post} and \textit{Meme Only}.}
\label{tab:multi_platform_results}
\end{table*}

We evaluate four representative multimodal models (LLaVA-v1.6-Vicuna-13B-hf, GLM-4.1V-9B-Thinking, Qwen3-VL-8B-Instruct, and Gemini-3) across three platforms (4chan, X, and Weibo) under two input settings: \textit{Meme + Post} and \textit{Meme Only}.

Overall, incorporating post text consistently improves performance across platforms (Table \ref{tab:language-statistics}), particularly in recall and F1, indicating the importance of complementary linguistic context. Among the evaluated models, Qwen3-VL-8B-Instruct achieves the strongest and most stable performance under both input settings, yielding the highest accuracy and F1 on all three platforms. GLM-4.1V-9B-Thinking also performs competitively, especially on X and Weibo, though it exhibits reduced recall in the meme-only setting.

In contrast, LLaVA-v1.6-Vicuna-13B-hf consistently underperforms, suggesting limited robustness to cross-platform and cross-lingual variation. Gemini-3 demonstrates high precision but notably low recall on X and Weibo, indicating a conservative prediction tendency that limits its effectiveness in multi-label hate detection.

These results highlight substantial performance variability across platforms and input modalities, underscoring the challenges posed by linguistic diversity, cultural context, and multimodal interactions in real-world hate speech detection.

\subsection{Case Study}

\begin{figure}[htbp]
\centering
\includegraphics[width=0.95\linewidth]{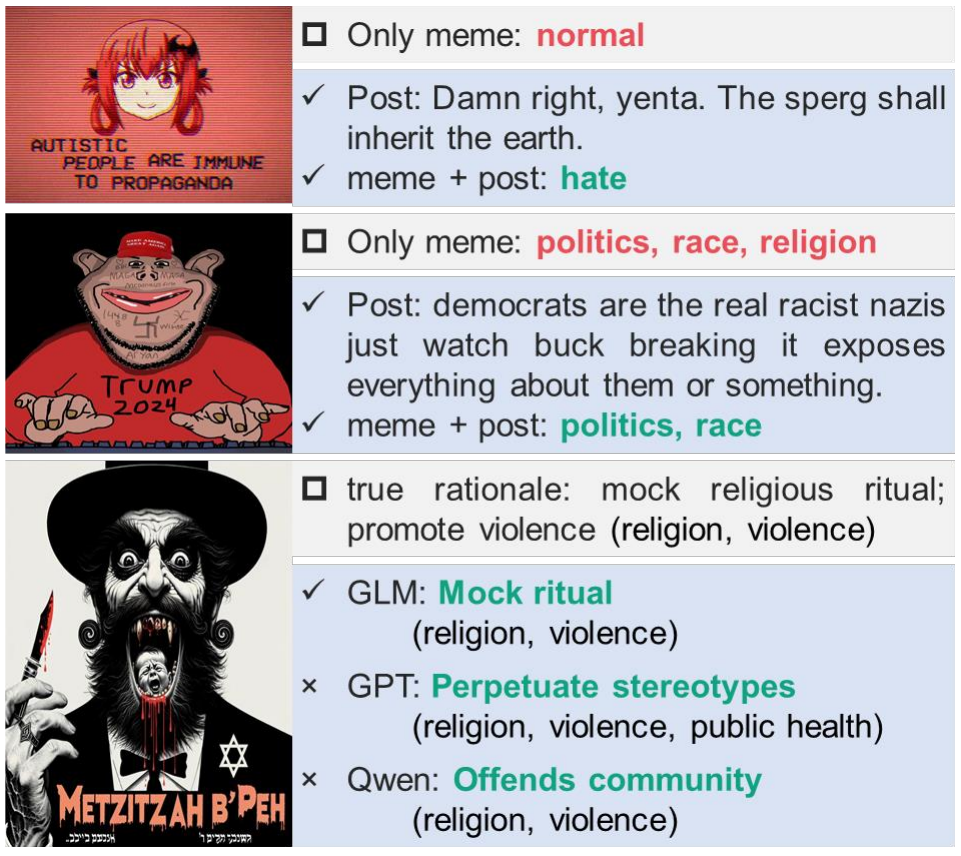}
\caption{Case study showing model predictions under meme-only and meme+post settings. Adding a post changes the prediction from normal to hate (top), and refines category predictions (middle). The bottom part compares model-generated rationales, where GLM-4.1V aligns better with ground truth than GPT-4o and Qwen2.5-VL.}
\label{fig:case}
\end{figure}

To qualitatively illustrate model behavior on M\textsuperscript{3}, we examine representative examples shown in Figure \ref{fig:case}, highlighting the impact of textual context and differences in rationale generation across models.

The case studies in Figure \ref{fig:case} demonstrate the critical role of accompanying textual context in meme understanding. In the first example, the meme image alone appears ambiguous and is classified as normal. However, when the post text is introduced, implicit hostility becomes explicit, leading the model to correctly revise its prediction to hate. This highlights how textual cues can surface latent intent that is not visually apparent.

In the second example, the inclusion of post text reduces model uncertainty by narrowing overgeneralized predictions. Without textual context, the model assigns multiple hate-related categories, reflecting ambiguity in visual interpretation. The post text provides additional constraints, enabling the model to refine its output to a smaller, more precise set of categories.

The bottom part of Figure \ref{fig:case} compares rationales generated by different models. GLM-4.1V-9B-Thinking produces rationales that are more semantically faithful to the ground truth, often using concise and well-aligned verb–object phrases. In contrast, GPT-4o and Qwen2.5-VL tend to generate broader or more generic rationales that capture the overall tone but miss specific discriminatory intent. These qualitative differences suggest that effective rationale generation favors semantic alignment over surface-level lexical overlap, particularly in multimodal hate analysis tasks.

\section{Related Work}
\label{sec:app-related-work}

Details regarding the size, annotation status, and language of unimodal hate speech datasets are presented in Table \ref{tab:Unimodal_datasets}, as supplemented in Section \ref{sec:related-work} of the paper.

\begin{table*}[t]
\centering
\renewcommand{\arraystretch}{1.2}
\resizebox{\textwidth}{!}{
\begin{tabular}{cccccc}
\toprule
\textbf{Dataset} & \textbf{Size} & \textbf{Label} & \textbf{Classification} & \textbf{Language} & \textbf{methods} \\ \midrule
Waseem's & 16,914 & Sexist, Racist, Neither & Multi-class & English & Manual \\ \midrule
Founta's & 80,000 & \makecell{Offensive, Abusive, Hate speech, \\ Aggressive, Cyberbullying, Spam, Normal} & Multi-class & English & Manual \\ \midrule
L-HSAB & 5846 & hate, abusive, normal & Multi-class & Arabic & Manual \\ \midrule
CHSD & 17,430 & hate, normal & binary & Chinese & Manual \\ \midrule
\multirow{4}{*}{OLID} & \multirow{4}{*}{14,100} & offensive, not offensive & \multirow{4}{*}{\makecell{Hierarchical \\multi-label}} & \multirow{4}{*}{English} & \multirow{4}{*}{manual} \\ \cmidrule(lr){3-3}
& & targeted insult, untargeted \\ \cmidrule(lr){3-3}
& & individual, group, other \\ \midrule
\multirow{3.5}{*}{HateXplain} & \multirow{3.5}{*}{20,148} & hate, offensive, normal & \multirow{3.5}{*}{\makecell{Hierarchical \\multi-label}} & \multirow{3.5}{*}{English} & \multirow{3.5}{*}{manual} \\ \cmidrule(lr){3-3}
& & \makecell{African, Islam, Jewish, Heterosexual, \\ Women, Refugee, Arab, Caucasian, Hispanic, Asian} \\ \midrule
\multirow{2.5}{*}{HATEDAY} & \multirow{2.5}{*}{240,000} & Hateful, Offensive, Neutral & \multirow{2.5}{*}{Hierarchical multi-label} & \multirow{2.5}{*}{\makecell{Arabic, English, \\ French, German, Indonesian, \\ Portuguese, Spanish, Turkish}} & \multirow{2.5}{*}{manual} \\ \cmidrule(lr){3-3}
& & Politics, National Origin, Gender, Religion, Sexual Orientation \\ \midrule
\multirow{5.5}{*}{HateBRXplain} & \multirow{5.5}{*}{7,000} & Offensive, Non-offensive & \multirow{5.5}{*}{Hierarchical multi-label} & \multirow{5.5}{*}{Portuguese} & \multirow{5.5}{*}{manual} \\ \cmidrule(lr){3-3}
& & highly, moderately, slightly \\ \cmidrule(lr){3-3}
& & \makecell{xenophobia, racism, homophobia, sexism, \\ religious intolerance, partyism, \\ apology for the dictatorship, antisemitism, fatphobia} \\
\bottomrule
\end{tabular}
}
\caption{\textbf{Summary of unimodal hate speech datasets.}
For OLID, \textit{targeted insult} and \textit{untargeted} are secondary labels under the primary label \textit{offensive}; further, \textit{individual}, \textit{group}, and \textit{other} are tertiary labels under \textit{targeted insult}. 
For HateXplain, \textit{hate}, \textit{offensive}, and \textit{normal} are primary labels, while secondary labels denote specific targeted communities, such as \textit{African}, \textit{Islam}, and others.
For HATEDAY, \textit{hateful}, \textit{offensive}, and \textit{neutral} are primary labels; \textit{politics}, \textit{national origin}, \textit{gender}, \textit{religion}, \textit{sexual orientation} are secondary labels under \textit{hateful}.
For HateBRXplain, \textit{highly}, \textit{moderately}, \textit{slightly} are secondary labels under the primary label \textit{offensive}; Tertiary labels such as \textit{xenophobia} are hate speech groups.
}
\label{tab:Unimodal_datasets}
\end{table*}

\end{document}